\documentclass[12pt]{article}
\usepackage{latexsym}
\usepackage{amssymb}
\usepackage{amsmath}
\usepackage{graphicx}

\newcommand{\C}{8\pi}

\newcommand{\bsq}{a^{2}(t)}
\newcommand{\bt}{a(t)}
\newcommand{\bdot}{\dot{a}(t)}
\newcommand{\bdotsq}{\dot{a}^{2}(t)}
\newcommand{\gprime}{g^{\prime}( r)}
\newcommand{\g}{g( r)}
\newcommand{\bdotdot}{\ddot{a}(t)}
\newcommand{\mew}{\mu(t,  r)}
\newcommand{\p}{p(t,  r)}
\newcommand{\alf}{\alpha(t,  r)}

\begin{document}

\title{Inhomogeneous cosmologies with tachyonic dust as dark matter}
\author {{\small A. Das \footnote{e-mail: das@sfu.ca}} \\
\it{\small Department of Mathematics} \\ \it{\small Simon Fraser
University, Burnaby, British Columbia, Canada V5A 1S6} \and {\small A. DeBenedictis
\footnote{e-mail: adebened@sfu.ca}} \\
\it{\small Department of Physics} \\
\it{\small Simon Fraser University, Burnaby, British Columbia,
Canada V5A 1S6}}
\date{{\small February 15, 2004}}
\maketitle

\begin{abstract}
\noindent A cosmology is considered driven by a stress-energy
tensor consisting of a perfect fluid, an inhomogeneous pressure
term (which we call a ``tachyonic dust'' for reasons which will
become apparent) and a cosmological constant. The inflationary,
radiation dominated and matter dominated eras are investigated in
detail. In all three eras, the tachyonic pressure decreases with
increasing radius of the universe and is thus minimal in the
matter dominated era. The gravitational effects of the dust,
however, may still strongly affect the universe at present time.
In case the tachyonic pressure is positive, it enhances the
overall matter {\em density} and is a candidate for dark matter.
In the case where the tachyonic pressure is negative, the recent
acceleration of the universe can be understood without the need
for a cosmological constant. The ordinary matter, however, has
positive energy density at all times. In a later section, the
extension to a variable cosmological term is investigated and a
specific model is put forward such that recent acceleration and
future re-collapse is possible.
\end{abstract}

\vspace{3mm}

\section{Introduction}
\qquad There are compelling reasons to study a cosmology which is
not homogeneous. Inhomogeneous models were studied early on by
Lema\^{i}tre \cite{ref:lemaitre} and Tolman \cite{ref:tolman} and
by many authors since. Misner \cite{ref:misner}, for example,
postulated a chaotic cosmology in which the universe began in a
highly irregular state but which becomes regular at late times.
The models presented here possess exactly this property, which
will be realized in a later section. That is, in the matter
phase, the deviation from FLRW spatial geometry is \emph{minimal}
and we show this by calculating the Gaussian curvature of two
spheres in all phases. The curvature of a two sphere is the same
for all values of $r$ in the matter domain yielding a
three-dimensional space which is isometric to a sphere. Our
present location may therefore be anywhere in this universe  and 
there is no conflict with
observational cosmology. The book by Krasi\'{n}ski also contains
many inhomogeneous models which do not require us to be located
at the center of symmetry \cite{ref:krasinskibook}. Some more
recent studies dealing with inhomogeneous cosmologies include
\cite{ref:fein}, \cite{ref:iban1}, \cite{ref:bark},
\cite{ref:bark2}, \cite{ref:iban2}, \cite{ref:karasinskipaper}
and \cite{ref:barm}.

\qquad Inhomogeneous cosmological models are not at odds with
astrophysical data. It is well known that inhomogeneities in the
early universe will generate anisotropies in the cosmic microwave
background radiation (CMB). Such effects have been studied by
many groups (\cite{ref:arnau1}, \cite{ref:saez},
\cite{ref:arnau2}, \cite{ref:fullana}) using density amplitudes
and sizes of inhomogeneities corresponding to those of observed
current objects (galactic clusters, the Great Attractor and
voids). These studies, utilizing a range of reasonable
parameters, have found that temperature fluctuations in the CMB,
$\Delta T/\langle T\rangle$, ($\langle T\rangle$ being the mean
temperature and $\Delta T$ the deviation from the mean) amount to
no more than about $10^{-7} - 10^{-5}$, which is compatible with
observation. Also, arguments to reconcile inhomogeneous solutions
with cosmological observations may be found in
\cite{ref:karasinskipaper}. The inhomogeneity referred to in this
paper is a ``radial'' inhomogeneity compatible with spherical
symmetry and therefore its effect on the CMB is potentially more
difficult to detect than the (small) angular deviations.

\qquad In general, at very high energies, our knowledge of the
state of the universe is highly limited and special assumptions
about the matter content and symmetry should be relaxed. It
therefore seems reasonable to investigate solutions which, at
least at early times, are less symmetric than the FLRW
scenarios.  A thorough exposition on various inhomogeneous
cosmological models may be found in the book by Krasi\'{n}ski
\cite{ref:krasinskibook}.

\qquad In section 2 we consider a cosmology consisting of two
fluids, a perfect fluid (motivated by the successful standard
cosmology) and ``tachyonic'' dust. We use the term tachyonic due
to the association of this source with space-like vectors in the
stress-energy tensor. This terminology is also popular in
string-theory motivated cosmologies commenced by the pioneering
works of Mazumdar, Panda, P\'{e}rez-Lornezana \cite{ref:mazumdar}
and Sen \cite{ref:sen1} and studied by many others (see, for
example, \cite{ref:tachcos1}, \cite{ref:tachcos2},
\cite{ref:feins}, \cite{ref:fks}, \cite{ref:tachcos3},
\cite{ref:matlock}, \cite{ref:tachcos4}, \cite{ref:tachcos6},
\cite{ref:tachcos8}, \cite{ref:tachcos5}, \cite{ref:tachcos7}
\cite{ref:tachcos9} and references therein). It should be pointed
out that in \emph{neither} the case presented here nor the string
theory motivated case is the source acausal as will be pointed
out below .

\qquad The tachyonic dust is chosen as a dark matter candidate for
several reasons. First, it provides one of the simplest
extensions to the standard perfect fluid cosmology and it is
hoped that this model will provide insight into more complex
scenarios. Second, as will be seen below, the tachyonic dust is a
source of pressure or tension without energy density and
cosmological observations strongly imply that there exists a
large pressure or tension component in our universe. This
pressure also affects the overall effective mass of the universe.
Multi-fluid models in the context of charged black holes in
cosmology have been studied in \cite{ref:daskay}

\qquad In section \ref{sec:varlam} we consider an extension of
the model to the case of variable cosmological term. We discuss
in detail how making this term dynamical affects the fate of the
universe.

\qquad Finally, this paper utilizes a number of techniques for
analyzing global properties of the manifold and it is hoped that
this will provide a useful reference for the mathematical
analysis of cosmological models.

\section{Tachyonic dust and perfect fluid universe}
\qquad We consider here a model of the universe which contains
both a perfect fluid and tachyonic dust. This source possesses
the desirable properties mentioned in the introduction. Namely,
the dust contribution is a source of pressure as is required for
the recent accelerating phase of the universe. A tachyonic dust is
the simplest model which contributes to pressure and it will be
shown that this pressure also makes a contribution to the mass of
the universe. This field is therefore also a potentially
interesting candidate for dark matter.

\qquad Aside from spherical symmetry, the sole assumption is that
the eigenvalues of stress-energy tensor be real. We may therefore
write
\begin{equation}
T^{\mu}_{\;\nu}=\left[\mew
+\p\right]u^{\mu}u_{\nu}+\p\delta^{\mu}_{\;\nu}+ \alf
w^{\mu}w_{\nu}, \label{eq:stresstens}
\end{equation}
with
\begin{equation}
u^{\beta}u_{\beta}=-1, \;\;\;\;\; w^{\beta}w_{\beta}=+1,
\;\;\;\;\; u^{\beta}w_{\beta}=0. \nonumber
\end{equation}
Here $\mew$, $\p$ and $\alf$ are the fluid energy density, fluid
pressure, and tachyonic {\em pressure} (or tension) respectively.
By comparison of the $\alf$ term in (\ref{eq:stresstens}) to the
stress-energy tensor of regular dust, it can be seen why we choose
the term ``tachyonic dust'' to describe this source. Notice that a
dust associated with a space-like vector possesses the desirable
property in that it yields solely a pressure. It will be shown
that this tension may produce the observed acceleration of the
universe at late times \cite{ref:riess}, \cite{ref:perl}. The
source is \emph{not} acausal as the algebraic structure of
(\ref{eq:stresstens}) is exactly similar to that of an
anisotropic fluid which is a causal source under minor 
restrictions and is often used in general relativity (see 
\cite{ref:makhark}, \cite{ref:ivanov}, \cite{ref:dev} and 
references therein). 

\qquad The time coordinate, $t$, may be chosen to be coincident
with the proper time along a fluid streamline (the comoving
condition). This gauge, along with spherical symmetry, allows a
special class of metrics to be written as
\begin{subequations}
\begin{align}
d\sigma^{2}:=&\left[\frac{d r^2}{1-\epsilon r^{2}+eg( r)} +
 r^2\,d\theta^{2} +  r^2\sin^{2}\theta\,d\phi^{2}\right], \label{eq:submetric} \\
ds^{2}=& -dt^{2}+ \bsq\, d\sigma^{2}. \label{eq:metric}
\end{align}
\end{subequations}
This form is particularly convenient as one may readily analyze
differences between models presented here and the standard FLRW
models (the $e\rightarrow 0$ limit). Therefore, $e$ may be
interpreted as the tachyon coupling constant. It is easy to show
that (\ref{eq:submetric}-\ref{eq:metric}) falls in the
Tolman-Bondi class of metrics, used extensively in studies of
inhomogeneous cosmologies.

\qquad Using (\ref{eq:stresstens}) and (\ref{eq:metric}) in the
Einstein equations with cosmological constant yields:
\begin{subequations}
\begin{align}
\C\mew+\Lambda=&3\left[\frac{\bdot}{\bt}\right]^{2}+
\frac{3\epsilon}{\bsq} -\frac{e}{ r^{2}\bsq}\left[
r\g\right]^{\prime}, \label{eq:einstone}
\\
\C
\p-\Lambda=&-2\frac{\bdotdot}{\bt}-\left[\frac{\bdot}{\bt}\right]^{2}
-\frac{\epsilon}{\bsq}+\frac{e\gprime}{2 r\bsq}, \label{eq:einsttwo} \\
\C\alf=&-\frac{e r}{2\bsq}\left[\frac{\g}{ r^{2}}
\right]^{\prime}, \label{eq:einstthree}
\end{align}
\end{subequations}
where dots represent partial derivatives with respect to $t$ and
primes with respect to $ r$.

\qquad Enforcing conservation on (\ref{eq:stresstens}) yields two
non-trivial equations:
\begin{subequations}
\begin{align}
\mew_{,t} +\frac{\bdot}{\bt}\left\{3\left[\mew+ \p\right]
+\alf\right\}&=0, \label{eq:cons1} \\
\left[\p+\alf\right]_{, r} +\frac{2\alf}{ r}&=0. \label{eq:cons2}
\end{align}
\end{subequations}
Throughout this paper, restrictions $\bt > 0$, $\bdot \neq 0$ and
$r>0$ are assumed in solving the differential equations. In case
the tachyon parameter $e=0$, one gets back the standard FLRW
cosmology.

\qquad The orthonormal Riemann components will be useful:
\begin{subequations}
\begin{align}
R_{\hat{t}\hat{ r}\hat{t}\hat{ r}}=& -\frac{\bdot}{\bt}=
R_{\hat{t}\hat{\theta}\hat{t}\hat{\theta}},
\label{eq:rtrtr} \\
R_{\hat{ r}\hat{\theta}\hat{ r}\hat{\theta}}\equiv R_{\hat{
r}\hat{\phi}\hat{ r}\hat{\phi}}=& \frac{1}{\bsq r}\left[
r\bdotsq+ \epsilon r
-\frac{e\gprime}{2} \right], \label{eq:rrthrth} \\
R_{\hat{\theta}\hat{\phi}\hat{\theta}\hat{\phi}}=&
\left[\frac{\bdot}{\bt}\right]^{2}+\frac{2\epsilon r -e \gprime
}{ r\bt^2}, \label{eq:rthphthph}
\end{align}
\end{subequations}
as well as those related by symmetry (hatted indices denote the
orthonormal frame). The solutions, being local and valid in some
domain, need not possess the neighbourhood near $r=0$.  The
singularity at $r=0$ will be addressed in a later section.

\qquad In cosmology, two measurable parameters considered are the
Hubble parameter $H(t)$ and the deceleration parameter, $q(t)$.
These are:
\begin{subequations}
\begin{align}
H(t):=&\frac{\bdot}{\bt}=-R_{\hat{t}\hat{ r}\hat{t}\hat{ r}}, \\
q(t):=& -\frac{\bt\bdotdot}{\bdotsq}.
\end{align}
\end{subequations}
The field equations, (\ref{eq:einstone}), (\ref{eq:einsttwo}) and
(\ref{eq:einstthree}) yield
\begin{subequations}
\begin{align}
\left[H(t)\right]^{2}=&\frac{\C}{3}\mew +\frac{\Lambda}{3}-
\frac{\epsilon}{\bsq}
+\frac{e\left[ r\g\right]^{\prime}}{3 r^{2}\bsq}, \label{eq:modeinstone} \\
6q(t)\left[H(t)\right]^{2}=&\C\left[\mew +3\p\right] -2\Lambda
-\frac{e r}{2\bsq} \left[\frac{\g}{ r^{2}}\right]^{\prime}, \label{eq:modeinsttwo} \\
\left[H(t)\right]^{2}\left[2q(t)-1\right] =&\C\p -\Lambda
+\frac{1}{\bsq}\left[\epsilon -\frac{e\gprime}{2 r}\right]
\label{eq:modeinstthree}.
\end{align}
\end{subequations}

\qquad To study inhomogeneity, the orthonormal Riemann components
of the three-dimensional sub-space (\ref{eq:submetric}) are
useful:
\begin{subequations}
\begin{align}
\tilde{R}_{\hat{ r}\hat{\theta}\hat{ r}\hat{\theta}}\equiv
\tilde{R}_{\hat{ r}\hat{\phi}\hat{ r}\hat{\phi}}
=& \epsilon-\frac{e\gprime}{2 r}, \\
\tilde{R}_{\hat{\theta}\hat{\phi}\hat{\theta}\hat{\phi}}=&
\epsilon-\frac{e\g}{ r^{2}}.
\end{align}
\end{subequations}
where the tilde is used to denote quantities calculated using the
three dimensional subspace metric of $t=t_{0}$ spatial
hyper-surfaces (\ref{eq:submetric}).

\qquad Finally, it is useful to define a measure of the
inhomogeneity of the spatial universe via an inhomogeneity
parameter:
\begin{subequations}
\begin{align}
I( r):=&\frac{\tilde{R}_{\hat{ r}\hat{\theta}\hat{
r}\hat{\theta}}}
{\tilde{R}_{\hat{\theta}\hat{\phi}\hat{\theta}\hat{\phi}}} \\
 =&\frac{ r\left[ 2\epsilon r-
e\gprime\right]}{2\left[\epsilon r^{2}-e\g\right]}.
\label{eq:inhomgen}
\end{align}
\end{subequations}
A homogeneous space is characterized by $\frac{dI(r)}{dr}\equiv
0$. Specifically, for the FLRW ($e=0$) limit, $I( r) \equiv 1$.

\qquad We next investigate the three major eras of cosmological
evolution.

\subsection{Matter dominated era}
\qquad In the universe's recent history, the galaxies which
constitute the bulk of the ordinary matter have negligible motion
relative to the cosmic expansion. Therefore the pressure of
ordinary matter is approximately zero. Reasonable physics also
demands that $\mew >0$. Setting the pressure equal to zero from
the equation (\ref{eq:einsttwo}) yields (assuming $\bdot\neq0$)
\begin{equation}
\frac{\left[\bt\bdotsq\right]^{\cdot}}{\bdot} -\Lambda\bsq=-
\left[\epsilon -\frac{e}{2 r}\gprime\right] = -C =\mbox{a
constant}. \label{eq:matspeeqn}
\end{equation}
Here, $C$ is the constant of separation. Solving this equation
for $\g$ one obtains
\begin{equation}
e\g=\left(\epsilon-C\right) r^{2}+eb, \label{eq:matg}
\end{equation}
with $b$ a constant arising from integration.

\qquad The equation for the expansion factor can be analyzed
using techniques, many of which are well known in cosmology. We
include details here for completion. The equation, after an
integration, may be written as:
\begin{equation}
\frac{1}{2}\bdotsq-\frac{M_{0}}{\bt}-
\frac{\Lambda}{6}\bt^{2}=-\frac{C}{2}. \label{eq:matenergy}
\end{equation}
Here $M_{0}$ is a constant arising from the integration. In the
standard cosmology this equation is often compared to total energy
conservation and similar equations have been studied at least as
early as Lemaitr\'{e} and Eddington \cite{ref:lem} \cite{ref:edd}.
The terms on the left hand side correspond to a kinetic energy,
gravitational potential energy and vacuum energy respectively.
The total energy being constant, $\left(-C/2\right)$. The
constant $M_{0}$ may therefore be interpreted as an effective mass
of the universe and it is of interest to investigate how the
tachyon affects this constant.

\qquad The equation (\ref{eq:matenergy}) may be used in
(\ref{eq:einstone}) along with (\ref{eq:matg}) to give the current
effective mass of the universe:
\begin{equation}
M_{0}=\mew\left[\frac{4}{3}\pi a^{3}(t)\right] +\frac{eb}{6
r^{2}}\bt. \label{eq:matmass}
\end{equation}
The second term in this equation gives the tachyonic contribution
to the effective mass of the universe and therefore represents
the present mass due to dark matter (which is independent of
$\Lambda$ in this section).

\qquad The fluid and tachyonic energy density and pressures are
given by:
\begin{subequations}
\begin{align}
8\pi\mew=& -\Lambda+3\left[\frac{\bdot}{\bt}\right]^{2}
+\left(3C-\frac{eb}{ r^{2}}\right) \frac{1}{\bsq},  \\
8\pi\p\equiv& 0, \\
8\pi\alf=&\frac{eb}{r^{2}\bsq}. \label{eq:matalpha}
\end{align}
\end{subequations}
Note that the tachyon pressure can be very small today (for large
$\bt$) although its effects through (\ref{eq:matmass}) can be
very large.

\qquad Acceleration in the matter phase may be analyzed by
studying the equations (\ref{eq:modeinsttwo}) and
(\ref{eq:modeinstthree}):
\begin{subequations}
\begin{align}
2q -1=& \frac{1}{\bdotsq} \left[C -\Lambda\bsq\right], \\
H^{2}(t)\left(2q-1\right)=&\frac{C}{\bt}-\Lambda , \\
q=&\mew \left[\frac{4}{3}\pi \frac{\bsq}{\bdotsq}\right]
+\frac{eb}{6 r^{2}\bdotsq}
-\frac{\Lambda}{3}\frac{\bsq}{\bdotsq}. \label{eq:matacpar}
\end{align}
\end{subequations}
Note that for positive $\mew$, $q$ may be negative 
even with
$\Lambda=0$. This result indicates that the tachyonic dust may
drive the relatively recent acceleration phase indicated by
supernova observations \cite{ref:riess}, \cite{ref:perl},
\cite{ref:riess2}. (Recall that positive acceleration corresponds
to a negative deceleration parameter.) We will discuss in 
a later section the values the parameters in 
(\ref{eq:matacpar}) must possess for this scenario. However, the 
emphasis in
this paper will be on Lambda driven acceleration, the tachyon
assuming the role of dark matter.

\qquad Acceleration can also be studied by differentiating
(\ref{eq:matenergy}) to obtain
\begin{equation}
\bdotdot=-\frac{M_{0}}{\bsq}+\frac{\Lambda}{3}\bt.
\label{eq:matacceleq}
\end{equation}
The above ``force'' equation nicely demonstrates the fact that
positive $M_{0}$ tends to produce an attractive force whereas
positive $\Lambda$ produces a negative or repulsive force. The
tachyonic effect is inherent in $M_{0}$ via (\ref{eq:matmass}).

\begin{figure}[ht!]
\begin{center}
\includegraphics[bb=87 373 427 696, clip, scale=1.0, keepaspectratio=true]{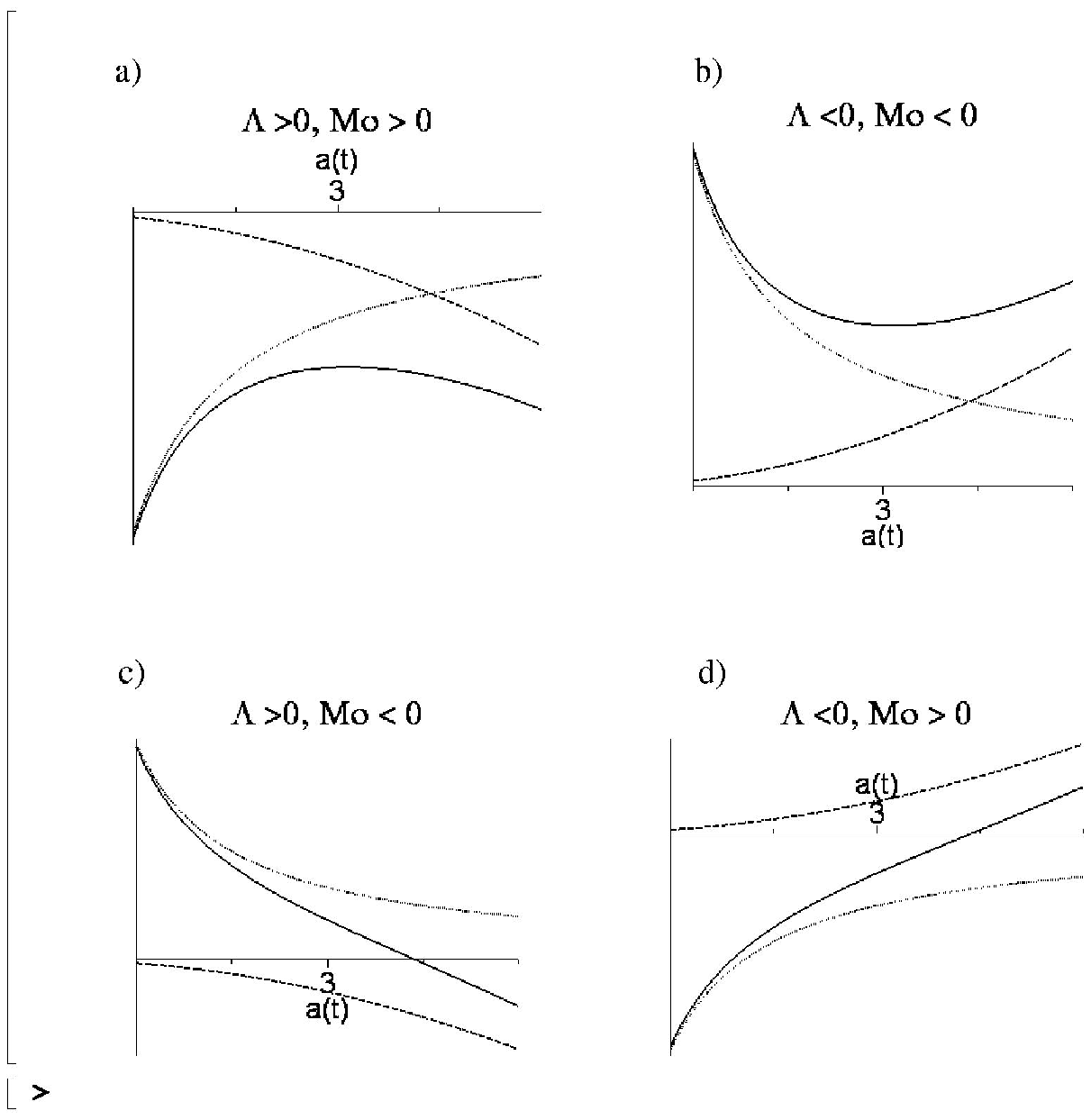}
\caption{{\small Effective potentials in the matter phase. Dashed
lines denote cosmological potential, $-\frac{\Lambda}{6\bsq}$,
grey lines denote matter potential, $-\frac{M_{0}}{\bt}$, and
solid black lines denote net effective potential. Re-collapse is
possible for scenarios (a), (b) and (d). Parameters to produce
the graphs are $|\Lambda|=0.1$ and $|M_{0}|=1$ although the
qualitative picture remains unchanged for other values.}}
\label{fig:two}
\end{center}
\end{figure}

\qquad The fate of the universe is governed by the scale factor,
$\bt$. In general, the equation for $\bt$ cannot be solved
explicitly. Here we use effective potential techniques to study
properties of $\bt$. Figure \ref{fig:two} shows plots of the
effective potentials due to the matter fields (grey line
indicating the function $-M_{0}/\bt$) and the cosmological term
(dashed line indicating the function $-\frac{\Lambda}{6\bsq}$) as
well as the sum of the two (solid) for various signs of $M_{0}$
and $\Lambda$.

\qquad From the figure it can be seen that for $C > 0$ situations
depicted in figures \ref{fig:two} (a) and (d) allow solutions
which re-collapse even for $\Lambda=0$. For $C < 0$, the
configurations in figures \ref{fig:two} (b) and (d) allow for
re-collapse (there are no re-collapse solutions for $C<0$ if
$\Lambda = 0$). In 2 (c) re-collapse is impossible.

\qquad It is of interest to study the geometry of spatial sections
generated by this solution. As mentioned in the introduction, the
space-like hyper-surfaces are not surfaces of constant curvature.
As well, if we consider the global picture, then the parameters
discussed can also affect the topology of the universe. Spatial
hyper-surfaces at $t=t_{0}$ possess the line element (equation
\ref{eq:submetric} except for a scale factor)
\begin{equation}
d\sigma^{2}= \frac{d r^2}{1-C r^{2}+eb} +  r^2\,d\theta^{2} +
 r^2\sin^{2}\theta\,d\phi^{2}. \label{eq:mat3dsub}
\end{equation}
Although (\ref{eq:mat3dsub}) bears a close resemblance to the
standard FLRW line element, they are not equivalent. The
orthonormal Riemann components for (\ref{eq:mat3dsub}) yield:
\begin{subequations}
\begin{align}
\tilde{R}_{\hat{ r}\hat{\theta}\hat{ r}\hat{\theta}}=&C, \label{eq:orthriemmat1}\\
\tilde{R}_{\hat{\theta}\hat{\phi}\hat{\theta}\hat{\phi}}=& C
-\frac{eb}{ r^{2}} \label{eq:orthriemmat2},
\end{align}
\end{subequations}
and therefore, for $e\neq0$, the three dimensional hyper-surfaces
are not of constant curvature. For small $e$ deviations are
minimal and for $e=0$, the hyper-surface is of constant curvature
$C$. The inhomogeneity parameter (\ref{eq:inhomgen}) is
calculated to be
\begin{equation}
I( r)=\frac{{C r}^{2}}{{C r}^{2}-eb}. \label{eq:matinhom}
\end{equation}

\qquad If we wish to treat the solutions as global, then the
spatial topology may be studied. The two dimensional sub-manifold
($\theta=\pi/2$) of the three-metric (\ref{eq:mat3dsub}) possesses
line element
\begin{equation}
d\sigma_{(2)}^{2}=\frac{d r^2}{1-C r^{2}+eb}+ r^{2}\,d\phi^{2}.
\end{equation}
Transforming to the arc-length parameter, $l$, along a $
r$-coordinate curve, one can obtain
\begin{equation}
d\sigma_{(2)}^{2}=\frac{\left[ r^{\prime}(l)\right]^{2}}{A-C
r^{2}(l)}\, d l^2+ r^{2}(l)\,d\phi^{2} =dl^{2}+
r^{2}(l)\,d\phi^{2}, \label{eq:twoarc}
\end{equation}
with $A:=1+eb\,$ and $\left[ r^{\prime}(l)\right]^{2}=A-C
r^{2}(l)$. Integrating for $ r(l)>0$, the following solutions are
derived:
\begin{equation}
 r(l)=\left\{
\begin{array}{lll}
\sqrt{\frac{A}{2C}}\sin\left[\sqrt{C}(l-l_{0})\right] & \mbox{ for }  & C > 0\; , \; A > 0 \\
\sqrt{\frac{A}{2|C|}} \sinh\left[\sqrt{|C|}\left(l-l_{0}\right)
\right] & \mbox{ for }&
C < 0  \; , \; A > 0 \\
\sqrt{\frac{A}{2}}\left(l-l_{0}\right) & \mbox{ for }& C = 0 \; ,
\; A > 0 \; ,
\end{array}
\right.
\end{equation}
($l_{0}$ is a constant arising from integration). It is clear
from (\ref{eq:twoarc}) that $l$ is a geodesic coordinate. From
the periodicity of the sine function, it may be seen that the two
conjugate points on the radial geodesic congruences are given by
$\, r(l_{0})= r(l_{0}+\pi/\sqrt{C}) =0$. Thus, one concludes that
spatially closed universes correspond only to $C > 0$.

\subsection{Radiation dominated era}
\qquad Here we study the next major phase in the evolution of the
universe. The radiation dominated phase is characterized by the
relativistic fluid equation of state $\mew=3\p$. Using this along
with (\ref{eq:einstone}) and (\ref{eq:einsttwo}) yields:
\begin{equation}
\frac{1}{2}\left[\bsq\right]^{\cdot\cdot} -\frac{2}{3}\Lambda\bsq=
-\epsilon+\frac{5e}{12 r^{7/5}}\left[ r^{2/5}\g\right]^{\prime}
=-C,
\end{equation}
again $C$ is a separation constant. The equation for $\g$ is
satisfied by
\begin{equation}
e\g=\left(\epsilon-C\right) r^{2}+\frac{eb}{ r^{2/5}}.
\label{eq:radg}
\end{equation}

\qquad Solving for the scale factor, $\bt$, one obtains:
\begin{equation}
\bsq=\left\{
\begin{array}{lll}
-Ct^{2}+\kappa_{1}t+\kappa_{2} & \mbox{ for }  & \Lambda = 0  \\
\kappa_{1}e^{2\sqrt{\frac{\Lambda}{3}}t} +
\kappa_{2}e^{-2\sqrt{\frac{\Lambda}{3}}t} +\frac{3C}{2\Lambda}&
\mbox{ for }&
\Lambda > 0  \\
\kappa_{1}\sin\left[2\sqrt{\frac{-\Lambda}{3}}t\right] +
\kappa_{2}\cos\left[2\sqrt{\frac{-\Lambda}{3}}t\right]+
\frac{3C}{2\Lambda} & \mbox{ for }& \Lambda < 0  \; .
\end{array}
\right.
\end{equation}
Here, $\kappa_{1}$ and $\kappa_{2}$ are arbitrary constants of
integration. However, the domain of $t$ and the signs of these
constants must respect $\bsq>0$.

\qquad The densities and pressures are given by
\begin{subequations}
\begin{align}
8\pi\mew=&-\Lambda +3\left[\frac{\bdot}{\bt}\right]^{2}
+\frac{3}{\bsq}\left[C-\frac{eb}{5 r^{12/5}} \right], \\
8\pi\p=&\frac{8\pi}{3}\mew, \\
8\pi\alf=&\frac{6eb}{5 r^{12/5}\bsq}.
\end{align}
\end{subequations}

\qquad The Hubble parameter is calculated to be:
\begin{equation}
H(t)=\left\{
\begin{array}{lll}
\frac{1}{2}\frac{\kappa_{1}-2Ct}{-Ct^{2}+\kappa_{1}t+\kappa_{2}} & \mbox{ for }  & \Lambda = 0  \\
\sqrt{\frac{\Lambda}{3}}\left[
\frac{\kappa_{1}\exp\left(2\sqrt{\frac{\Lambda}{3}}t\right) -
\kappa_{2}\exp\left(-2\sqrt{\frac{\Lambda}{3}}t\right)}{\kappa_{1}e^{2\sqrt{\frac{\Lambda}{3}}t}
+ \kappa_{2}e^{-2\sqrt{\frac{\Lambda}{3}}t}+3C/\Lambda} \right] &
\mbox{ for }&
\Lambda > 0  \\
\sqrt{\frac{-\Lambda}{3}}
\left[\frac{\kappa_{1}\cos\left(2\sqrt{\frac{-\Lambda}{3}}t\right)
- \kappa_{2}\sin\left(2\sqrt{\frac{-\Lambda}{3}}t\right)}
{\kappa_{1}\sin\left(2\sqrt{\frac{-\Lambda}{3}}t\right) +
\kappa_{2}\cos\left(2\sqrt{\frac{-\Lambda}{3}}t\right)+3C/\Lambda}\right]
& \mbox{ for }& \Lambda < 0  \; .
\end{array}
\right.
\end{equation}

The deceleration parameter is provided by:
\begin{equation}
q(t)=\left\{
\begin{array}{lll}
\frac{4C\kappa_{2}+\kappa_{1}^{2}}{\left(\kappa_{1}-2Ct\right)^{2}}
& \mbox{ for }
& \Lambda = 0  \nonumber \\
-\frac{6\kappa_{1}\kappa_{2}\Lambda + 3C\left[ \kappa_{1}
 \exp\left(2\sqrt{\frac{\Lambda}{3}}t\right) + \kappa_{2}
\exp\left(-2\sqrt{\frac{\Lambda}{3}}t\right)\right]
+\Lambda\kappa_{1}^{2}\exp\left(4\frac{\Lambda}{3}t\right)
+\Lambda\kappa_{2}^{2}\exp\left(-4\frac{\Lambda}{3}t\right)}{\Lambda
\left[\kappa_{1}e^{2\sqrt{\frac{\Lambda}{3}}t} -
\kappa_{2}e^{-2\sqrt{\frac{\Lambda}{3}}t}\right]^{2}} & \mbox{ for
}&
\Lambda > 0  \nonumber \\
\frac{\kappa_{1}^{2}+\kappa_{2}^{2} + \left[
\kappa_{1}\sin\left(2\sqrt{\frac{-\Lambda}{3}}t\right) +
\kappa_{2}\cos\left(2\sqrt{\frac{-\Lambda}{3}}t\right)
\right]^{2} +\left(\frac{3C}{\Lambda}\right) \left[
\kappa_{1}\sin\left(2\sqrt{\frac{-\Lambda}{3}}t\right) +
\kappa_{2}\cos\left(2\sqrt{\frac{-\Lambda}{3}}t\right) \right]}
{\left[\kappa_{1}\sin\left(2\sqrt{\frac{-\Lambda}{3}}t\right) +
\kappa_{2}\cos\left(2\sqrt{\frac{-\Lambda}{3}}t\right)\right]^{2}}
& \mbox{ for }& \Lambda < 0 \; . \nonumber
\end{array}
\right. \nonumber
\end{equation}

\qquad Finally, the inhomogeneity parameter of the spatial
hyper-surfaces is given by:
\begin{equation}
I( r)=\frac{\left( 5\,{ C r}^{{\frac {12}{5}}}+eb
\right)}{5\left( {C r}^{{\frac {12}{5}}}-eb \right)}.
\end{equation}

\qquad The presence of the tachyon affects the spatial geometry.
Here spatial geometry is again studied via the arc-length
parameter $l$. The geodesic equation along an $r$-coordinate
curve yields
\begin{subequations}
\begin{align}
\left[\frac{d r(l)}{dl}\right]^{2} + 2V( r(l)) =1
\label{eq:radspatgeod} \\
V( r):=\frac{1}{2}C r^{2} -\frac{eb}{2 r^{2/5}}.
\end{align}
\end{subequations}
One may analyze (\ref{eq:radspatgeod}) via similar ``effective
potential'' techniques as in the dynamics. For positive $C$, $
r(l)$ is bounded regardless of the sign of the tachyonic
potential, $eb$ (spatially closed universe). For negative $C$, all
allowed solutions are unbounded (spatially open universe). For
$C=0$, the spatial universe is also open.

\subsection{Inflationary era}
\qquad We now investigate the inflationary phase. It is generally
believed that the universe experienced tremendous expansion over
a short period of time. There are many physical reasons for
believing in this scenario and an excellent review may be found
in \cite{ref:brandeninfl}. Some studies of the scalar tachyon's
relevance to inflation may be found in \cite{ref:fks},
\cite{ref:tachcos3}, \cite{ref:tachcos4}, \cite{ref:tachcos6}. In
the scenario presented here, the tachyon does not play the role
of the inflaton. However, the inflationary phase provides one
possible mechanism for the transition from high tachyon
concentration to low concentration.

\qquad Inflationary scenarios are generally supported by the
equation of state $\mew +\p=0$. This linear combination of
(\ref{eq:einstone}) and (\ref{eq:einsttwo}) yields:
\begin{equation}
\bsq\left[\ln|\bt|\right]^{\cdot\cdot}=\epsilon-\frac{e}{4 r^{3}}
\left[ r^{2}\g\right]^{\prime} = C.
\end{equation}
The solution for $\g$ is given by
\begin{equation}
eg( r)= \left(\epsilon-C\right) r^{2} + \frac{eb}{ r^{2}}\,.
\label{eq:infg}
\end{equation}
As well, the following modes are found for $\bt$:
\begin{subequations}
\begin{align}
\bt=&\beta_{0} e^{H t}& \mbox{for } C=0,\;\;\; \beta_{0}>0, \label{eq:infb1} \\
\bt=&\sqrt{\frac{C}{\beta_{1}}}\cosh\left[\sqrt{\beta_{1}}(t-t_{0})\right]
&\mbox{for } C>0,\;\;\; \beta_{1}>0, \label{eq:infb2} \\
\bt=&\sqrt{\frac{|C|}{\beta_{2}}}\sinh\left[\sqrt{\beta_{2}}(t-t_{0})\right]
&\mbox{for } C<0,\;\;\; \beta_{2}>0, \label{eq:infb5} \\
\bt=&\sqrt{\frac{C}{\beta_{2}}}
\sin\left[\sqrt{|\beta_{2}|}(t-t_{0})\right]
& \mbox{for } C<0,\;\;\; \beta_{2}<0, \label{eq:infb4} \\
\bt=&\sqrt{|C|}(t-t_{0}) &\mbox{for } C<0,\;\;\; \beta_{2}=0.
\label{eq:infb3}
\end{align}
\end{subequations}
Here $\beta_{0}$, $\beta_{1}$, $\beta_{2}$ and $H$ are constants
of integration.

\qquad The Hubble factor is given by
\begin{equation}
H(t)=\left\{
\begin{array}{lllll}
H=\mbox{constant} & \mbox{ for }  &  C = 0, \;\;\; \beta_{0}>0  \\
\sqrt{\beta_{1}}\tanh\left[\sqrt{\beta_{1}}(t-t_{0})\right] &
\mbox{ for }
&  C > 0, \;\;\; \beta_{1}>0  \\
\sqrt{\beta_{2}}\coth\left[\sqrt{\beta_{2}}(t-t_{0})\right] &
 \mbox{ for }  &  C < 0, \;\;\; \beta_{2} >0  \\
\sqrt{|\beta_{2}|}\cot\left[\sqrt{|\beta_{2}|}(t-t_{0})\right] &
 \mbox{ for }  &  C < 0, \;\;\; \beta_{2} < 0  \\
(t-t_{0})^{-1} & \mbox{ for }  &  C < 0, \;\;\; \beta_{2}=0 .
\end{array}
\right.
\end{equation}
The corresponding deceleration parameter
\begin{equation}
q(t)=\left\{
\begin{array}{lllll}
-1 & \mbox{ for }  &  C = 0, \;\;\; \beta_{0}>0  \\
-\coth^{2}\left[\sqrt{\beta_{1}}(t-t_{0})\right] & \mbox{ for }  &
  C > 0, \;\;\; \beta_{1}>0  \\
-\tanh^{2}\left[\sqrt{\beta_{2}}(t-t_{0})\right] & \mbox{ for }  &
 C < 0, \;\;\; \beta_{2} >0  \\
\tan^{2}\left[\sqrt{|\beta_{2}|}(t-t_{0})\right] & \mbox{ for }  &
  C < 0, \;\;\; \beta_{2} < 0  \\
0 & \mbox{ for }  &  C < 0, \;\;\; \beta_{2}=0\, .
\end{array}
\right.
\end{equation}

\qquad The source terms are:
\begin{subequations}
\begin{align}
8\pi\mew=-8\pi\p =&\frac{1}{\bsq}\left[3C+\bdotsq
+\frac{eb}{ r^{4}}\right] -\Lambda, \\
8\pi\alf=&\frac{2eb}{ r^{4}\bsq}
\end{align}
\end{subequations}
from which it can be seen that the tachyon is naturally diluted
by the presence of a scale factor which increases rapidly. The
fluid density and pressures, however, need not dilute as their
expressions contain terms proportional to $\bdot/\bt$ which may
tend to constant (as in (\ref{eq:infb1}) and (\ref{eq:infb2})) or
increase (as in (\ref{eq:infb4})). We demonstrate several scenarios next.

\begin{figure}[ht!]
\begin{center}
\includegraphics[bb=72 155 480 739, clip, scale=0.5, keepaspectratio=true]{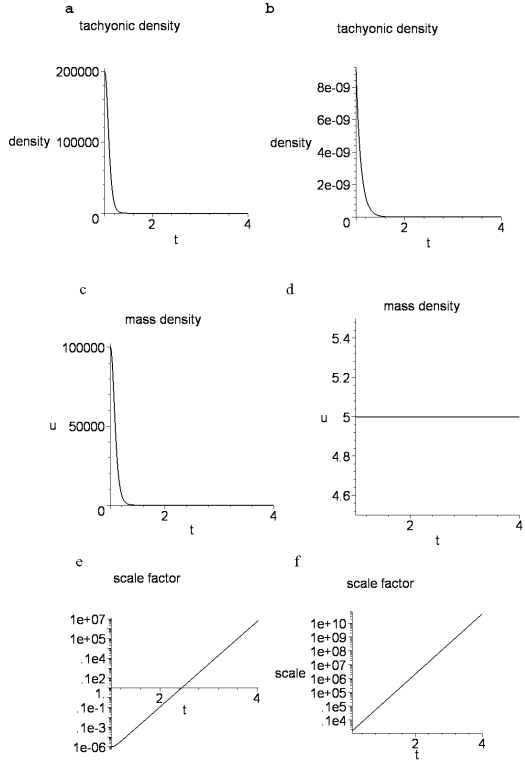}
\caption{{\small Inflationary scenarios: graphs on the left represent a $C > 0$ model whereas graphs on the right represent a $C=0$ model. Space-time coordinates are measured in $10^{-24}$m and densities are scaled accordingly. Graphs a) and b) represent the evolution of $\alf$, graphs c) and d) the evolution of $\mew$ and graphs e) and f) the increase in the scale factor (see text for explanation).}} \label{fig:infex}
\end{center}
\end{figure}

\qquad In the figure \ref{fig:infex}.
The graphs on the left represent the scenario with $C>0$ (``closed inflation'') whereas the graphs on the right represent the $C = 0$ scenario (``flat inflation'') at some fixed value of $r$. Both scenarios are with $\Lambda=0$ so that $\mew$ represents the energy density of all fields (dominated by the inflaton, with minor contributions from other fields) save for the tachyon, whose density is given by $\alf$ in graphs a) and b). The space-time coordinates possess units of $10^{-24}$ metres here. Note that for an acceptable interval of inflation (approx a few times $10^{-32}$s), we have, in the $C=0$ scenario, a dramatic decrease in the density of the tachyon field but not the necessarily the inflaton field. In this scenario, inflation must terminate by the standard phase transition of the inflaton field. At the end of inflation, the tachyon density is much smaller than the densities of the other matter which will be dominated by radiation leading to the radiation era. In the $C>0$ scenario, both $\alf$ \emph{and} $\mew$ vary with time although $\mew$ (initially primarily the inflaton) approaches a constant value while $\alf$ decreases as $a^{-2}(t)$ (this is not obvious from graph c, however it can easily be seen, by examining the analytic expressions for $\alf$ and $\mew$ with $C >0$, that $\mew$ possesses a term which does not decay with time whereas $\alf$ does not possess such a term). It is a simple matter to show that parameters exist to produce an increase in the expansion factor by many orders of magnitude. The figures \ref{fig:infex} show this although their time axes have been truncated to show the behaviour of $a(t)$ more clearly.

\qquad As inflation progresses, both models yield a tachyon density whose value decreases to a smaller value than $\mew$. This value can be made small enough as not to intertfere with the physical processes that must have occured during the radiation dominated era.

\qquad The spatial geometry is again studied using the arc-length
parameter, $l$, as in the matter dominated era. In this case
\begin{equation}
 r(l)=\left\{
\begin{array}{lll}
\frac{1}{\sqrt{2C}}\left\{\left(1+4Ceb\right)^{1/2}
\sin\left[2\sqrt{C}\left(l-l_{0}\right) \right]
+1\right\}^{1/2} & \mbox{ for }  & C > 0  \\
\frac{1}{\sqrt{2|C|}}  \left\{\left(4|C|eb-1\right)^{1/2}
\sinh\left[2\sqrt{|C|}\left(l-l_{0}\right) \right]
-1\right\}^{1/2} & \mbox{ for }&
C < 0  \\
\sqrt{\left(l-l_{0}\right)^{2}-eb} & \mbox{ for }& C = 0  \; .
\end{array}
\right.
\end{equation}
Here we see that, from periodicity of the sine function, $C>0$
again can yield the closed spatial universe.

Finally, the inhomogeneity parameter in this phase is
\begin{equation}
I( r)={\frac {C{ r}^{4}+eb}{C{ r}^{4}-eb}}.
\end{equation}

\section{An extension to variable Lambda cosmology}
\label{sec:varlam} \qquad Recent experiments suggest that the
universe is presently in an accelerating phase. If one accepts
that the net mass of the universe is positive, then the present
acceleration can be explained by the figure 2a alone. Thus, the
choice $\Lambda > 0$ must be made. In case $\Lambda >0$ is a
constant, re-collapse is incompatible with acceleration.
Therefore, we consider the generalization of the previous
sections to the variable $\Lambda(t)$ case. This scenario has
relevance in light of recent models (mainly based on supergravity
considerations) which predict that the dark energy decreases and
that the universe re-collapses within a time comparable to the
present age of the universe (see \cite{ref:kalinde} and
references therein).

\qquad Time dependent fields with equation of state $p(t)\approx
- \mu(t)$ have been employed in the literature to explain certain
evolutionary periods requiring positive acceleration. There are
also compelling reasons from particle physics for treating the
cosmological term as a dynamic quantity (see \cite{ref:varlam1},
\cite{ref:varlam2}, \cite{ref:varlam3} \cite{ref:varlam4} and
references therein).

\qquad The field equations (\ref{eq:einstone}),
(\ref{eq:einsttwo}), (\ref{eq:einstthree}) formally remain the
same with the exception that $\Lambda = \Lambda(t)$. However, the
conservation equation (\ref{eq:cons1}) needs to be augmented by
an additional term. The definitions of the matter, radiation and
inflationary phases are retained exactly as before. Therefore,
the equations for $\g$ in all three phases remain intact.

\qquad The solutions for $\g$ can be summarized as:
\begin{equation}
e\g=(\epsilon-C) r^{2} +\frac{eb}{ r^{\nu}},
\end{equation}
with
\begin{equation}
\nu=\left\{
\begin{array}{lll}
0 & \mbox{ for the matter phase} \\
2/5 & \mbox{ for the radiation phase } \\
2 & \mbox{ for the inflationary phase }.
\end{array}
\right.
\end{equation}
The three-geometries are specified as
\begin{equation}
d\sigma^{2}=\frac{d r^{2}}{1-C r^{2}+\frac{eb}{ r^{\nu}}} +
r^{2}\,d\theta^{2} +  r^{2}\sin^{2}\theta \, d\phi^{2},
\end{equation}
\begin{subequations}
\begin{align}
\tilde{R}_{\hat{ r}\hat{\theta}\hat{ r}\hat{\theta}} =&C
+\frac{\nu eb}{2 r^{\nu+2}},\label{eq:genorthriem1} \\
\tilde{R}_{\hat{\theta}\hat{\phi}\hat{\theta}\hat{\phi}} =&C
-\frac{ eb}{2 r^{\nu+2}}, \label{eq:genorthriem2}
\end{align}
\end{subequations}
from which one obtains
\begin{equation}
I( r)=\frac{2C r^{\nu+2}+\nu eb}{2\left[C r^{\nu+2} -eb\right]} .
\end{equation}

\qquad The field equations are
\begin{subequations}
\begin{align}
8\pi\mew + \Lambda(t)&=\frac{1}{\bsq}\left\{3 \left[\bdotsq
+C\right] + \frac{\left(\nu-1\right)eb}{ r^{\nu+2}}\right\},  \label{eq:varlam1} \\
8\pi\p -\Lambda(t) &=-\frac{1}{\bsq} \left\{2\bt\bdotdot +\bdotsq
+C +\frac{\nu eb}{2 r^{\nu+2}} \right\}. \label{eq:varlam2} \\
8\pi\alf &=\frac{\left(\nu+2\right)eb}{2 r^{\nu+2}\bsq}
\label{eq:varlam3},
\end{align}
\end{subequations}
with conservation laws:
\begin{subequations}
\begin{align}
\mew_{,t}+\frac{\dot{\Lambda}(t)}{8\pi} +H(t)
\left\{3\left[\mew+\p\right] +\alf\right\}&=0, \\
\left[\p+\alf\right]_{,r} +\frac{\left(\nu+2\right)eb}{8\pi
r^{\nu+3}\bsq}&=0.
\end{align}
\end{subequations}

\qquad The dynamical quantities are given by
\begin{equation}
\bdotdot=-\left\{\frac{4\pi}{3} \left[\mew +\alf +3\p\right] \bt
\right\} +\frac{1}{3}\Lambda(t)\bt . \label{eq:varlamadotdot}
\end{equation}
The Hubble parameter and the deceleration parameter are furnished
as
\begin{subequations}
\begin{align}
H^{2}(t)=&\frac{8\pi}{3}\mew +\frac{\Lambda(t)}{3} -\frac{C}{\bsq}
-\frac{(\nu-1)eb}{3 r^{\nu+2}\bsq} , \\
H^{2}(t)\left[2q(t)-1\right] =&8\pi\p -\Lambda(t)
+\frac{1}{\bsq}\left[C+ \frac{\nu eb}{2 r^{\nu+2}}\right].
\end{align}
\end{subequations}

\qquad As far as experimental evidences are concerned, the matter
domain is the most relevant. Therefore, the maximum information
possible will be elicited from the field equations for that
domain. Setting $\p= \nu = 0$ and integrating (\ref{eq:varlam2})
yields the ``energy'' conservation equation:
\begin{equation}
\frac{1}{2}\bdotsq -\frac{1}{6\bt}\int_{t_{2}}^{t} \Lambda(\tau)
\,d[a^{3}(\tau)] -\frac{M_{0}}{\bt} = -\frac{C}{2}
\label{eq:varlamenergy}.
\end{equation}
Again $M_{0}$ arises from the integration and represents the
total effective mass of the universe. Furthermore, $t_{2}$ is
another constant representing the beginning of the matter phase.
In terms of the matter fields, the mass is:
\begin{subequations}
\begin{align}
M_{0}=&\mew\left[\frac{4}{3}\pi a^{3}(t)\right] +\frac{eb}{6
r^{2}}\bt +\frac{a^{3}(t)}{6} \Lambda(t)
-\frac{1}{6}\int_{t_{0}}^{t}
\Lambda(\tau)\,d\left[a^{3}(\tau)\right], \label{eq:varlammass} \\
=&\left[\mu(t_{2},r)+\alpha(t_{2},r)\right] \left[\frac{4}{3}\pi
a^{3}(t_{2})\right] +\frac{a^{3}(t_{2})}{6}\Lambda(t_{2}) .
\end{align}
\end{subequations}
Here, a possible interpretation is that the first term represents
the total mass of observed matter (``normal'' matter), the second
term the tachyonic contribution to the dark matter (non-baryonic
mass for pressure, ``dark energy'' for tension) and the third
gives rise to potential ``dark energy'' responsible for
acceleration.

\qquad The acceleration is provided by (\ref{eq:varlamadotdot})
and (\ref{eq:varlamenergy}) as:
\begin{eqnarray}
\bdotdot=&-\frac{M_{0}}{\bsq} +\frac{1}{2}\Lambda(t)\bt -
\frac{1}{6\bsq}\int_{t_{2}}^{t}\Lambda(\tau)
\,d\left[a^{3}(\tau)\right] \nonumber \\
=&-\left\{ \frac{\left[\mew+\alf\right] \left[\frac{4}{3}\pi
a^{3}(t) \right]}{\bsq}\right\} +\frac{1}{3}\Lambda(t)\bt .
\end{eqnarray}
In case $\mew > 0$, $eb > 0$ and $\Lambda(t) > 0$, the above
terms on the right hand side produce a combination of both
attractive and repulsive forces.

\qquad The Hubble parameter in the matter domain is provided by:
\begin{subequations}
\begin{align}
H^{2}(t)=&\frac{8}{3}\pi\left[\mew +\alf\right]
+\frac{\Lambda(t)}{3}-\frac{C}{\bsq} \label{eq:hsqvarlam}
 \\
=&\frac{2M_{0}}{a^{3}(t)}-\frac{C}{\bsq} +\frac{1}{3 a^{3}(t)}
\int_{t_{2}}^{t}\Lambda(\tau)\, d\left[a^{3}(\tau)\right] .
\end{align}
\end{subequations}
The deceleration parameter in this domain is
\begin{equation}
H^{2}(t)\left[2q(t)-1\right]= \frac{C}{\bsq} -\Lambda(t).
\label{eq:varlammatdec}
\end{equation}

\qquad It is clear from (\ref{eq:varlammatdec}) that $q$ can be
positive, negative or zero depending on the values of $C$ and
$\Lambda(t)$. A specific model will be proposed which accommodates
a spatially closed, re-collapsing universe with an accelerating
period in the matter domain. In the cosmology presented here,
this may be realized by setting $\Lambda(t_{2})> 0$, $C >0$ and
$eb >0$. Observations indicate that $C$ has a value very close to
zero. A $C\leq 0$ universe has cubic divergent volume at all times
save the origin when the volume is zero. If, however, $C$ is
extremely small yet positive, one has finite large volume in the
matter domain without contradicting observations.

\qquad The time periods for inflation, radiation and matter
dominated eras are $[\epsilon, t_{1}]$, $[t_{1},t_{2}]$ and
$[t_{2},T/2]$ respectively. The time $T/2$ indicates the
initiation of re-collapse and thus represents the half-life of
the universe. Of course, the boundaries separating the domains
are not sharp as we have indicated and therefore the above simply
represents a rough guideline.

\begin{figure}[ht]
\begin{center}
\includegraphics[bb=51 245 541 587, clip, scale=0.5, keepaspectratio=true]{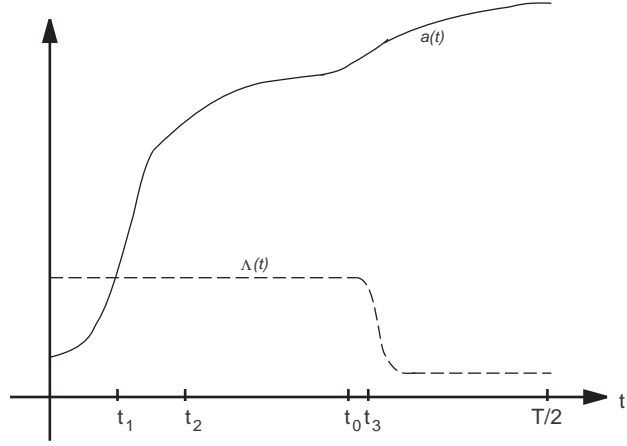}
\caption{{\small A possible scenario for the evolution of the
universe. The present time is denoted by $t=t_{0}$ and the
half-life of the universe denoted by $t=T/2$. The solid line
represents the qualitative evolution of the scale factor and the
dashed line the cosmological term. }} \label{fig:evolution}
\end{center}
\end{figure}

\qquad A possible evolutionary scenario is depicted in figure
\ref{fig:evolution}. Here we plot both the scale factor $\bt$ and
cosmological term $\Lambda(t)$ as a function of cosmic time. The
scale factor increases greatly during the inflationary phase (in
the model presented in the figure, the inflation is driven by
some matter field, not the cosmological term). This is followed
by a decelerating phase and, near the present time, a period of
acceleration follows. This scenario is based on the tachyonic
positive pressure model and therefore this acceleration is
$\Lambda$ driven. To allow for re-collapse, the cosmological term
decays (starting at $t=t_{3}$) so that $\bdotdot$ becomes negative
causing deceleration and eventual re-collapse. The figure is
symmetric about $T/2$. Furthermore, one may have a cyclic universe
where the scenario repeats after the ``big crunch''.

\qquad A suitable $\Lambda(t)$ function may be defined by
\begin{equation}
\Lambda(t)=\left\{
\begin{array}{llll}
\Lambda_{0}\;\;\;\mbox{a positive constant} & \mbox{ for }  & 0 \leq t < t_{1}  \\
\Lambda_{0} & \mbox{ for }& t_{1} \leq t < t_{2}  \\
\Lambda_{0} & \mbox{ for }& t_{2} \leq t < t_{3} \\
\Lambda_{0} -\epsilon_{2}(t-t_{3})^{2} + \epsilon_{4}(t-t_{4})^{4}
& \mbox{ for }& t_{3} < t < T/2 \; ,
\end{array}
\right.
\end{equation}
with $\epsilon_{2}>0$ and $\epsilon_{4}>0$.

\qquad The expansion factor for this case is given by:
\begin{equation}
 a(t)=\left\{
\begin{array}{lll}
\sqrt{\frac{C}{\beta_{0}}} \cosh(\sqrt{\beta_{0}}t) \;\; & \mbox{ for } 0 \leq t < t_{1}  \\
\sqrt{\frac{3C}{2\Lambda_{0}}+\kappa_{1}
\exp\left[2\sqrt{\frac{\Lambda_{0}}{3}}t\right] + \kappa_{2}
\exp\left[-2\sqrt{\frac{\Lambda_{0}}{3}}t\right]}
\;\; & \mbox{ for } t_{1} \leq t < t_{2} \label{eq:avarlam} \\
f^{-1}(t-t_{2})  \;\;& \mbox{ for } t_{2} \leq t < t_{3} \; ,
\end{array}
\right.
\end{equation}
$f(a):= \int_{a(t_{2})}^{a(t)} \frac{dx}
{\sqrt{\frac{\Lambda_{0}}{3}x^{2} +\left[2M_{0}
-\frac{\Lambda_{0}}{3}a^{3}(t_{2})\right]x^{-1}-C}}$. There are
enough arbitrary parameters in (\ref{eq:avarlam}) so that $\bt$
can be joined smoothly in the three phases if one wishes to
enforce sharp boundaries between the phases.

\qquad The function $a(t)$ satisfies the formidable
integro-differential equation
\begin{equation}
\left[\bdot\right]^{2} -\frac{\Lambda_{0}}{3}\bsq
-\frac{2M_{0}-\Lambda_{0} a^{3}(t)}{\bt} +\frac{1}{3\bt}
\int_{t_{3}}^{t}\left(\epsilon_{2}\tau^{2}- \epsilon_{4} \tau^{4}
\right)f\left[a^{3}(\tau)\right] =0
\end{equation}
in the interval $t_{3} \leq t < T/2$. The above equation is too
difficult to solve analytically at this stage.

\qquad The spatial geometry for $C > 0$ is governed by $ r(l)$ as
\begin{equation}
d\sigma^{2}=dl^{2}+ r^{2}(l) \left(d\theta^{2}+\sin^{2}\theta\,
d\phi^{2} \right),
\end{equation}
\begin{equation}
 r(l)=\left\{
\begin{array}{lll}
\frac{1}{\sqrt{2C}}\left\{\left(1+4Ceb\right)^{1/2}
\sin\left[2\sqrt{C}\left(l-l_{0}\right) \right]
+1\right\}^{1/2} & \mbox{ for inflation}  \\
F^{-1}\left(l-l_{0}\right) & \mbox{ for radiation}  \\
\sqrt{\frac{1+eb}{2C}} \sin\left[\sqrt{C}\left(l-l_{0}\right)
\right] & \mbox{ for matter} \; .
\end{array}
\right.
\end{equation}
Here, $F( r):=\int\frac{d r}{\sqrt{1-C r^{2}+eb/ r^{2/5}}}\:$. By
previous discussions, in all phases the physical universes are
closed. Moreover, the total volume corresponding to the three
dimensional spatial sub-manifold in the matter phase is given by
\begin{equation}
\frac{2\pi\left(1+eb\right)}{C}
\left[a(t_{0})\right]^{3}\int_{l_{0}}^{l_{0}+\pi/\sqrt{C}}
\sin^{2}\left[\sqrt{C}\left(l-l_{0}\right)\right]\, dl
=\frac{\pi^{2}\left(1+eb\right)}{C^{3/2}}\left[a^{3}(t_{0})\right],
\end{equation}
(Note that in the limit $C\rightarrow 0_{+}$, the above volume
diverges).

\begin{figure}[ht!]
\begin{center}
\includegraphics[bb=2 95 586 761, clip, scale=0.5, keepaspectratio=true]{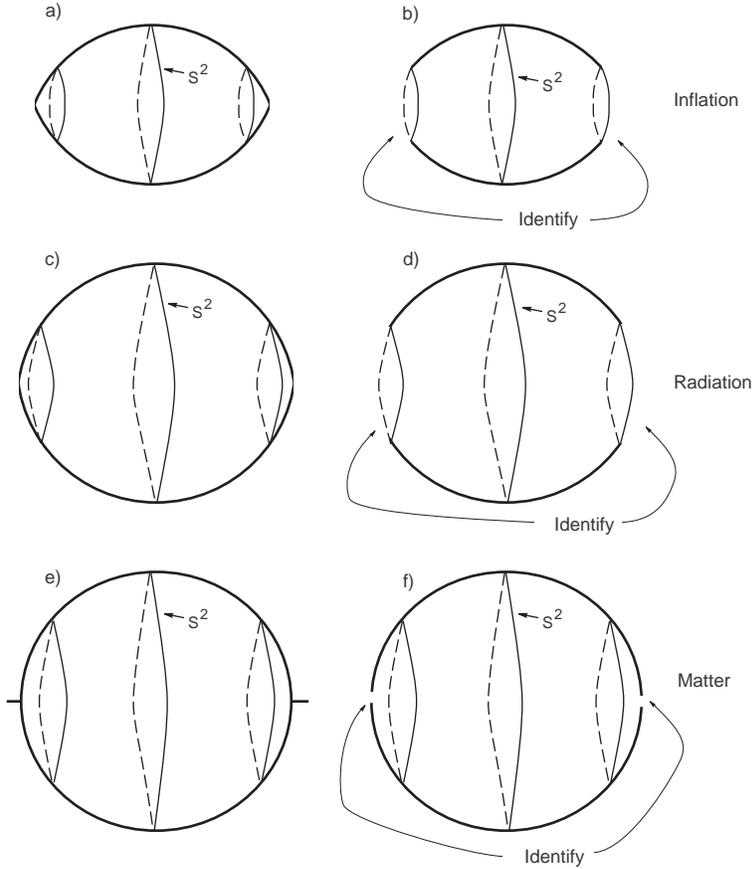}
\caption{{\small Possible global pictures for spatial sections in
the inflationary (a, b), radiation (c, d) and matter (e, f)
dominated eras. The diagrams on the left include the points
corresponding to $r=0$. The diagrams on the right have a
neighborhood about $r=0$ removed and the boundaries identified.
Note that in the matter domain the surfaces are isometric to
spheres yet singularities still exist at $r=0$ as indicated by
the ``hairs'' in diagram e.}} \label{fig:globals}
\end{center}
\end{figure}

\qquad We now wish to address the singularity at $r=0$. The two
dimensional geometries (restricted to $\theta=\pi/2$) yield:
\begin{equation}
d\sigma_{2}^{2}=\frac{dr^{2}}{1-C r^{2}+\frac{eb}{r^{\nu}}}
+r^{2}\,d\phi^{2}.
\end{equation}
These two-dimensional surfaces embedded in a three-dimensional
Euclidean space possess the following Gaussian curvatures:
\begin{equation}
 K(r,\phi)=\left\{
\begin{array}{lll}
C+\frac{eb}{r^{4}} \;\; > 0 & \mbox{ for inflation}  \\
C+ \frac{eb}{5r^{12/5}}\;\; > 0 & \mbox{ for radiation}  \\
C \;\; > 0 & \mbox{ for matter} \; .
\end{array}
\right.
\end{equation}
In the matter domain, the surface is \emph{locally isometric to a
spherical surface of radius} $1/\sqrt{C}$. However, the original
three dimensional spaces in the equation (\ref{eq:genorthriem2})
all exhibit a singularity at the limit $r\rightarrow 0_{+}$.
Therefore, some possible global pictures for these three
dimensional spaces are provided in figure \ref{fig:globals}.

\qquad In the figure \ref{fig:globals}, one of the angles is
suppressed so that latitudinal lines represent two-spheres. Two
possible scenarios exist; the figures on the left represent the
spatial manifold for inflation, radiation dominated and matter
dominated eras which include $r=0$ (the left and right points in
each figure). The figures on the right have the domains in the
neighborhood of $r=0$ excised. The left and right boundaries are
therefore identified. Note that as the evolution progresses, the
anisotropy of the spatial sections diminishes yielding a sphere
in the matter domain. This is therefore quite compatible with
observation. The poles of the sphere, however, are singular or
must be excised. The singularity appears to be ``soft'' in that
it is of the conical type. Also, the equations, being local,
are valid in a domain $r_{1} < r < r_{2}$ which need not include
$r=0$. It is likely that such a singularity would be absent in a
quantum theory of gravity which would be manifest at high
energies.

\section{Compatibility with current observations}
\qquad Current observations indicate a universe which is
approximately 5\% baryonic matter, 20\% non-baryonic matter and
75\% ``dark energy'' which is responsible for the recent
acceleration phase. The ``directly'' measurable quantities in
cosmology are $\mew$, $H$ and $q_{0}$. Roughly, in the present
epoch (and, as we are dealing with an inhomogeneous universe, in
our neighborhood of the universe) these quantities possess the
following approximate values:
\begin{subequations}
\begin{align}
\mu(t_{0}, r_{0}) \approx& 1.6\times 10^{-56}\, \mbox{m}^{-2}, \label{eq:nummu} \\
H^{2}:=&\left[\frac{\bdot}{\bt}\right]^{2} \approx 7.3\times
10^{-53}\,
\mbox{m}^{-2}, \label{eq:nuH} \\
q_{0} \approx& -0.4\;\;\;. \label{eq:numq}
\end{align}
\end{subequations}
Here $t_{0}$ and $ r_{0}$ are the current time and position
respectively.

\qquad The deceleration equation (\ref{eq:matacpar}) provides a
relationship (using the above parameters along with
(\ref{eq:matalpha})) between $\Lambda$ and $\alf$ (we assume that
any time variation in $\Lambda$ can be ignored):
\begin{equation}
\alpha(t_{0}, r_{0}) = \left( 8.0\times 10^{-2} \Lambda
-6.9\times 10^{-54}\right) \mbox{m}^{-2}\;. \label{eq:alphalambda}
\end{equation}
If there is no cosmological constant, then the second term in
this equation indicates the approximate value the tachyon tension
must possess in our region of the universe to drive the observed
acceleration.

\qquad If, on the other hand, the tachyon possesses positive
pressure (contributing all or in part to the non-baryonic dark
matter of the universe) then the acceleration is $\Lambda$ driven.
In such a case $\alpha(t_{0}, r_{0})$ may take on the following
values:
\begin{equation}
0 \leq \alpha(t_{0}, r_{0}) \lesssim 4 \mu(t_{0}, r_{0}) \approx
6.4\times 10^{-56} \, \mbox{m}^{-2}.
\end{equation}
The upper limit comes from noting the observational evidence that
the dark matter contribution is approximately four times the
baryonic contribution to the matter content. This sets a
restriction on the cosmological constant to be of the order
\begin{equation}
\Lambda = \mathcal{O}\left(10^{-52}\right)m^{-2} .
\end{equation}

\qquad Alternately we may begin the analysis by using equation
(\ref{eq:varlammatdec}) and solving for $\Lambda$ (with the
parameters quoted above)
\begin{equation}
\Lambda_{0}=\frac{C}{\bsq}- H^{2}\left(2q-1\right)
=\frac{C}{\bsq} +13.14\times 10^{-53}. \label{eq:lambdanot}
\end{equation}
Also, by equation (\ref{eq:hsqvarlam}), we may write
\begin{eqnarray}
3H^{2}&=&8\pi\left[\mew+\alf\right] +\Lambda_{0} -\frac{3C}{\bsq} \nonumber \\
2.19\times10^{-52}&=&8\pi\left[\mew+\alf\right] + 1.31\times
10^{-52}-2\frac{C}{\bsq} \label{eq:threehsq}
\end{eqnarray}
(in the last equation (\ref{eq:lambdanot}) has been used.)
Isolating the $\mew+\alf$ term and using in (\ref{eq:varlammass})
yields
\begin{equation}
\frac{M_{0}}{\frac{4\pi}{3}a^{3}(t_{0})} = 3.48\times10^{-54}
+\frac{1}{8\pi} \left[\frac{C}{a^{2}(t_{0})}+1.3\times10^{-52}
\right]\left[\frac{a(t_{2})}{a(t_{0})}\right]^{3} +\frac{C}{4\pi
a^{2}(t_{0})}. \label{eq:univdens}
\end{equation}
The $C/\bsq$ terms represent the present ``radius''-squared of the
universe. The left hand side of (\ref{eq:univdens}) is an
analogue of the present Newtonian density of the universe. The
above equation is therefore useful in determining the radius of
the universe given the density or vice-versa.

\section{Concluding remarks}
\qquad This paper considers a simple cosmological model
consisting of perfect fluid matter supplemented with a ``tachyonic
dust''. The perfect fluid, with positive mass density, makes up
the ordinary matter as in the standard cosmology. The tachyonic
dust term is a source of pressure which, interestingly, can
increase the effective mass of the universe. In this case it
could potentially be utilised as a source of dark matter although
the clustering properties need to be studied. In case the
tachyonic dust term is a source of tension, it may be responsible
for the observed recent acceleration of the universe. This model 
provides the simplest pressure enhancing extension to the 
successful FLRW scenario. At late times, the solution generates a 
geometry compatible with FLRW.

\section*{Acknowledgements}
\qquad The authors thank their home institutions for various
support during the production of this work. Also, A. DeB. thanks
the S.F.U. Mathematics department for kind hospitality. A. Das
thanks Dr. S. Kloster for useful informal discussions. We thank the anonymous referees for helpful suggestions. 
\newpage
\linespread{0.6}
\bibliographystyle{unsrt}

\begin{thebibliography}{10}
{\small

\bibitem{ref:lemaitre}
G.~Lema\^{i}tre,
\newblock {\em Ann. Soc. Sci. Bruxelles} {\bf A53} (1933) 51. Engl.
trans. {\em Gen. Rel. Grav.} {\bf 29} (1997) 641.

\bibitem{ref:tolman}
R.~C.~Tolman,
\newblock {\em Proc. Nat. Acad. Sci. USA} {\bf 20} (1934) 169. Reprint
{\em Gen. Rel. Grav.} {\bf 29} (1997) 935.

\bibitem{ref:misner}
C.~Misner,
\newblock {\em Astrophys. J.} {\bf 151} (1968) 431.

\bibitem{ref:krasinskibook}
A.~Krasi\'{n}ski,
\newblock {\em Inhomogeneous Cosmological Models} (Cambridge University
Press, Cambridge, 1997).

\bibitem{ref:fein}
A.~Feinstein, J.~Ib\'{a}\~{n}ez and P.~Labraga,
\newblock {\em J. Math. Phys.} {\bf 36} (1995) 4962.

\bibitem{ref:iban1}
J.~Ib\'{a}\~{n}ez and I.~Olasagasti,
\newblock {\em J. Math. Phys.} {\bf 37} (1996) 6283.

\bibitem{ref:bark}
J.~D.~Barrow and K.~E.~Kunze,
\newblock {\em gr-qc/9611007}.

\bibitem{ref:bark2}
J.~D.~Barrow and K.~E.~Kunze,
\newblock {\em Phys. Rev.} {\bf D56} (1997) 741.

\bibitem{ref:iban2}
J.~Ib\'{a}\~{n}ez and I.~Olasagasti,
\newblock {\em Class. Quant. Grav.} {\bf 15} (1998) 1937.

\bibitem{ref:karasinskipaper}
A.~Krasi\'{n}ski,
\newblock {\em Proceedings of the 49th Yamada Conference on black holes and high-energy
astrophysics} Kyoto, Japan, (Universal Academic Press, Tokyo,
1998)

\bibitem{ref:barm}
J.~D.~Barrow and R.~Maartens,
\newblock {\em Phys. Rev.} {\bf D59} (1999) 043502.

\bibitem{ref:arnau1}
J.~V.~Arnau, M.~Fullana, L.~Monreal and D.~S\'{a}ez,
\newblock {\em Astroph. J.} {\bf 402} (1993) 359.

\bibitem{ref:saez}
D.~S\'{a}ez, J.~V.~Arnau and M.~Fullana,
\newblock {\em Mon. Not. Roy. Astr. Soc.} {\bf 263} (1993) 681.

\bibitem{ref:arnau2}
J.~V.~Arnau, M.~Fullana and D.~S\'{a}ez,
\newblock {\em Mon. Not. Roy. Astr. Soc.} {\bf 268} (1994) L17.

\bibitem{ref:fullana}
M.~Fullana, J.~V.~Arnau, and D.~S\'{a}ez,
\newblock {\em Mon. Not. Roy. Astr. Soc.} {\bf 280} (1996) 1181.

\bibitem{ref:mazumdar}
A.~Mazumdar, S.~Panda, A.~P\'{e}rez-Lorenzana,
\newblock {\em Nucl. Phys.} {\bf B614} (2001) 101.

\bibitem{ref:sen1}
A.~Sen,
\newblock {\em J. H. E. P.} {\bf 0204} (2002) 048.

\bibitem{ref:tachcos1}
G.~W.~Gibbons,
\newblock {\em Phys. Let.} {\bf B537} (2002) 1.

\bibitem{ref:tachcos2}
T.~Padmanabhan,
\newblock {\em Phys. Rev.} {\bf D66} (2002) 021301.

\bibitem{ref:feins}
A.~Feinstein,
\newblock {Phys. Rev.} {\bf D 66} (2002) 063511.

\bibitem{ref:fks}
A.~Frolov, L.~Kofman and A.~Starobinsky,
\newblock {\em Phys. Lett.} {\bf B545} (2002) 8.

\bibitem{ref:tachcos3}
L.~Kofman and A.~Linde,
\newblock {\em J. H. E. P.} {\bf 0207} (2002) 004.

\bibitem{ref:matlock}
P.~Matlock, R.~C.~Rashkov, K.~S.~Viswanathan and Y.~Yang,
\newblock {\em Phys. Rev.} {\bf D66} (2002) 026004.

\bibitem{ref:tachcos4}
M.~Sami, P.~Chingangbam and T.~Qureshi,
\newblock {\em Phys. Rev.} {\bf D66} (2002) 043530.

\bibitem{ref:tachcos6}
X.~Li, D.~Liu and J.~Hao,
\newblock {\em hep-th/0207146}.

\bibitem{ref:tachcos8}
J.~S.~Bagla, H.~K.~Jassal and T.~Padmanabhan,
\newblock {\em astro-ph/0212198}.

\bibitem{ref:tachcos5}
D.~Choudhury, D.~Ghoshal, D.~P.~Jatkar and S.~Panda,
\newblock {\em Phys. Let.} {\bf B544} (2003) 231.

\bibitem{ref:tachcos7}
C.~Kim, H.~B.~Kim and Y.~Kim,
\newblock {\em Phys. Let.} {\bf B552} (2003) 111.

\bibitem{ref:tachcos9}
G.~W.~Gibbons,
\newblock {\em hep-th/0301117}.

\bibitem{ref:daskay}
A.~Das and D.~Kay,
\newblock {\em Can. J. Phys.} {\bf 66} (1988) 1031.

\bibitem{ref:riess}
A.~G.~Reiss, et al,
\newblock {\em Astron. J.} {\bf 116} (1998) 1009.

\bibitem{ref:perl}
S.~Perlmutter, et al,
\newblock {\em Astrophys. J.} {\bf 517} (1999) 565.

\bibitem{ref:makhark}
M.~K.~Mak and T.~Harko,
\newblock {\em Proc. Roy. Soc. Lond.} {\bf A459} (2003) 393.

\bibitem{ref:ivanov}
B.~V.~Ivanov,
\newblock {\em Phys. Rev.} {\bf D65} (2002) 104011.

\bibitem{ref:dev}
K.~Dev and M.~Gleiser,
\newblock {\em Gen. Rel. Grav.} {\bf 34} (2002) 1793.

\bibitem{ref:riess2}
A.~G.~Reiss, et al,
\newblock {\em Astrophys. J.} {\bf 560} (2001) 49.

\bibitem{ref:lem}
G.~Lemaitr\'e,
\newblock {\em Ann. Soc. Sci. Brux.} {\bf A47} (1927) 49.

\bibitem{ref:edd}
A.~S.~Eddington,
\newblock {\em Mon. Not. Roy. Astron. Soc.} {\bf 90} (1930) 668.

\bibitem{ref:brandeninfl}
R.~Brandenberger,
\newblock {\em Proceedings of the International School on
Cosmology} Kish Island, Iran, (Kluwer, Dordrecht, 2000).

\bibitem{ref:kalinde}
R.~Kallosh and A.~Linde,
\newblock {\em J. Cosmol. Astropart. Phys.} {\bf 02} (2003) 002.

\bibitem{ref:varlam1}
P.~G.~Bergmann,
\newblock {\em Int. J. Theor. Phys.} {\bf 1} (1968) 25.

\bibitem{ref:varlam2}
R.~V.~Wagoner,
\newblock {\em Phys. Rev.} {\bf D1} (1970) 3209.

\bibitem{ref:varlam3}
A.~D.~Linde,
\newblock {\em JETP Let.} {\bf 19} (1974) 183.

\bibitem{ref:varlam4}
J.~M.~Overduin and F.~I.~Cooperstock,
\newblock {\em Phys. Rev.} {\bf D58} (1998) 043506.
 }

\end{thebibliography}

\end{document}